\documentclass[12pt]{article}

\usepackage{sbc-template}
\usepackage[T1]{fontenc}
\usepackage[utf8]{inputenc}
\usepackage{graphicx}
\usepackage{url}
\usepackage{natbib}
\usepackage{hyperref}

\usepackage{amsmath}
\usepackage{algorithm}
\usepackage[noend]{algpseudocode}
\algdef{SE}[EVENT]{Event}{EndEvent}[1]{\textbf{upon event}\ #1\
\algorithmicdo}{\algorithmicend\ \textbf{event}}%
\algtext*{EndEvent}

\newcommand{\mytitle}{Quality   of   Service    of   an   Asynchronous
  Crash-Recovery Leader Election Algorithm}

\hypersetup{
     pdftitle={\mytitle}, 
     pdfauthor={Vinícius A. Reis, Gustavo M. D. Vieira},
     pdfdisplaydoctitle=true,
     hidelinks
}

\bibpunct{[}{]}{,}{a}{}{,}
\renewcommand{\cite}[2][]{\citep[#1]{#2}}

\title{\mytitle}

\author{Vinícius A. Reis\inst{1} \and Gustavo M.  D.  Vieira\inst{1} }

\address{DComp -- CCGT -- UFSCar\\
         Sorocaba, São Paulo, Brasil
         \email{angiolucci@gmail.com, gdvieira@ufscar.br}}

\begin{document}

\maketitle  



\begin{abstract}
  In asynchronous distributed systems it is very hard to assess if one
  of the processes taking part in a computation is operating correctly
  or has failed. To overcome  this problem, distributed algorithms are
  created  using  unreliable  failure  detectors that  capture  in  an
  abstract way  timing assumptions  necessary to assess  the operating
  status of  a process. One particular  type of failure detector  is a
  leader  election,  that indicates  a  single  process that  has  not
  failed. The unreliability of these failure detectors means that they
  can make mistakes, however if they  are to be used in practice there
  must be  limits to the  eventual behavior of these  detectors. These
  limits are defined  as the \emph{quality of  service (QoS)} provided
  by the  detector. Many  works have tackled  the problem  of creating
  failure  detectors with  predictable  QoS, but  only for  crash-stop
  processes and synchronous systems.  This paper presents and analyzes
  the behavior of a new leader  election algorithm named NFD-L for the
  asynchronous crash-recovery failure model that is efficient in terms
  of its use of stable memory and message exchanges.
\end{abstract}

\section{Introduction}

Fault-tolerant distributed  systems are created by  the aggregation of
many non  fault-tolerant computer systems in  clusters, coordinated by
fault-tolerant  software.  These  commodity  clusters  are modeled  as
\emph{asynchronous} systems where  there are no bounds  to the message
transmission delays and processing time.  As a consequence, it is very
hard to assess if one of the processes taking part of a computation is
operating correctly or  has failed. To overcome  this limitation, many
distributed  algorithms assume  the  stronger  computational model  of
asynchronous    processes    augmented   with    unreliable    failure
detectors~\cite{chandra96}.  More than just flagging failed processes,
unreliable  failures  detectors  capture  in an  abstract  way  timing
assumptions  necessary to  the correct  operation of  many distributed
algorithms~\cite{lamport98,  chandra96}.  The unreliability  of  these
detectors is key to this abstraction:  errors can be made and failures
are  detected   eventually,  in   a  way   that  reflects   the  timing
uncertainties intrinsic to asynchronous distributed systems.

Unreliable as they are, failure detectors are enough to design correct
distributed algorithms~\cite{chandra96c}.  However, if  they are to be
used in  practice there  must be  limits to  the eventual  behavior of
these  detectors.   These limits  are  properties  of the  distributed
system  (processes  and  network   links),  of  the  failure  detector
algorithm  used  and  of  its  operational  parameters,  defining  the
\emph{quality     of     service     (QoS)}    provided     by     the
detector~\cite{chen02}.   The QoS  of a  failure detector  can have  a
direct  impact on  the performance  of a  distributed algorithm.   For
example,   the   Paxos   algorithm~\cite{lamport98}  uses   a   single
coordinator process as the sequencer that orders and distributes a set
of totally-ordered messages among a  group of processes. The selection
of the coordinator  is made by a failure detector  and the progress of
the  algorithm is  halted while  the coordinator  is failed.   If this
coordinator  is \emph{wrongly}  assumed to  be failed  by the  failure
detector,  Paxos  throughput  is  affected, even  if  the  mistake  is
eventually corrected.   The more mistakes the  failure detector makes,
the worse Paxos  real world performance will be. Thus,  QoS of failure
detectors is an important parameter to be considered.

Many works have tackled the problem of creating failure detectors with
predictable QoS. The seminal work of Chen et al.~\cite{chen02} defined
a set  of QoS metrics, created  a new failure detector  with a precise
model of  its QoS and used  this model to configure  its parameters to
match a desired  QoS. Other works have experimentally  studied the QoS
of failure detectors~\cite{falai05, nunes04},  created more robust QoS
models~\cite{sotoma06}  and proposed  application-specific  QoS for  a
system-wide failure detector  service~\cite{hayashibara04}.  All these
algorithms and QoS models were created assuming the crash-stop process
abstraction,  where  processes can  only  fail  by crashing  and  once
crashed they  never return to  the computation. The only  exception is
the  work of  Ma et.   al.~\cite{ma010}  that tackles  the problem  of
analyzing the  QoS of Chen's  algorithm in the  crash-recovery process
abstraction,  where  processes  fail  by crashing  but  later  recover
preserving their stable memory.

The  works regarding  QoS  of failure  detectors  also share  another,
rather  ironical, property:  many assume  a \emph{synchronous}  system
where there is a (possibly  unknown) bound to the message transmission
delays and  processing times.  In  this model  it is easier  to create
models  of  QoS,  usually  based  on  the  existence  of  synchronized
clocks. Despite being a reasonable assumption in current systems, take
for example the use of NTP in clusters, these algorithms are burdening
the system designer with a more stringent synchrony model than the one
required  by  the algorithm  itself.   One  exception  is one  of  the
algorithms proposed by  Chen et. al.~\cite{chen02}, which  in its turn
has weaker QoS model based on  the accuracy of a \emph{predictor}. For
asynchronous systems, the ability of the failure detector to ascertain
the state of a process based on the history of its past communications
is very important for its QoS, as shown by~\cite{nunes04} and explored
by~\cite{hayashibara04}. Thus, there is no robust QoS model for
asynchronous systems beyond  the model originally proposed  by Chen et
al..

Moreover, no previous  work has analyzed the QoS of  a leader election
algorithm.   A leader  election is  a  type of  failure detector  that
instead of indicating  if a single process has failed,  it indicates a
single process that  has \emph{not} failed, with the  identity of this
process agreed by all correct processes.  In some senses, this type of
failure detector  consumes less  system resources because  it monitors
only a  single process~\cite{larrea00}  and it is  the weaker  form of
failure  detector   required  to   solve  consensus~\cite{chandra96c}.
Paxos,  a   consensus  algorithm,  requires  a   leader  election  for
asynchronous systems composed by  crash-recovery processes.  There are
suitable  leader  election   algorithms  that  support  crash-recovery
processes~\cite{martin09},  but  unfortunately   there  are  no  known
algorithms with these properties that  have a suitable model for their
QoS.

This paper presents a new  leader election  for the asynchronous
crash-recovery failure model that is efficient  in terms of its use of
stable memory and  message exchanges.  Our algorithm named  NFD-L is an
extension of  Chen's algorithm and  retains its QoS properties  in the
absence of failures.  We analyze the QoS of this algorithm and present
experimental data  as a first step  to providing a complete  QoS model
for this failure detector in the presence of failures and recoveries.

\section{Preliminaries} 

In this section we will describe the basic concepts of distributed
systems and the relationship between its abstractions. We then use
those abstractions to present more sophisticated concepts, such as
failure models and failure detectors.

\subsection{Synchrony and Failure Models}

\emph{Asynchronous}  distributed systems  have no  bounds for  (i) how
much time it takes for a message to be delivered and (ii) how much time
it takes for a process to do  some computation.  \emph{Synchronous} distributed systems can rely on
hard  bounds for  message delay  and computing  time.  \emph{Partially
  synchronous}  distributed system  are  systems  where processes  and
links  behaves most  of the  time  asynchronously, but  there is  an
unknow    time    in   the    future    after    which   it    behaves
synchronously~\cite{cachin2011}.

Another property that defines a distributed system is process failure
models. The \emph{crash-stop} model considers \emph{correct} a process
that   never  crashes.    Once  crashed,   a  process   is  considered
\emph{faulty}  and  it  never recovers.   The  \emph{crash-recovery}
failure  model  considers correct  a  process  that never  crashes  or
crashes  and  recovers  a  finite  number  of  times.   Thus,  in  the
\textit{crash-recovery}  model  a faulty  process  is  a process  that
crashes  and never  recovers or  a process  that crashes  and recovers
infinitely~\cite{cachin2011}.

In  this  paper  we  consider asynchronous  processes  augmented  with
unreliable failure detectors.  This is a type of partially synchronous
distributed system  where the timing  assumptions are confined  to the
failure  detector.   We  also  assume  crash-recovery  processes  that
communicate with each other by  exchanging messages through links. The
links  may  drop  or  delay  messages,  but  only  deliver  a  message
previously sent by some process.

\subsection{Failure Detection and Leader Election}\label{sec_fd}

Given a distributed system composed by processes and links, a failure
detector  is   a  component  that  outputs   not  necessarily  correct
information about which processes are correct or faulty.
It works as a local component, queried by processes 
 in order to know about which processes are still correct. When a
process $q$ queries its local failure detector about the state of
process $p$, the only two possible responses it can receive  are either
\emph{trust}, if the failure detector trusts process $p$ is correct
or \emph{suspect}, if the failure detector suspects $p$ is faulty.
Failure  detectors  are  usually implemented  by  exchanging  messages
through  its  links~\cite{chen02}: a  process  $p$  sends a  heartbeat
message  to  another process  $q$  every  $\eta$  time units,  if  $q$
receives no  messages from $p$  after a  timeout plus a  safety margin
$\alpha$, the failure detector in $q$ will start suspecting $p$ may be
crashed.

An  important  result  is  that  it  is  possible  to  build  reliable
distributed systems on top of an unreliable failure
detector~\cite{chandra96}. It means a failure detector is not supposed
to be correct while the system behaves asynchronously~\cite{flp85}, it
may make mistakes  by suspecting correct processes  or trusting faulty
ones. However, assuming a partially synchronous system, eventually the
processes and  links will  behave synchronously  and then  the failure
detector  will  stop making  mistakes.   Based  on these  assumptions,
reliable distributed systems  are designed to be  \emph{safe} when the
system behaves asynchronously and only \emph{progress} when it behaves
synchronously~\cite{guerraoui00}.  Despite the mistakes it can make, a
failure  detector  is  a   powerful  abstraction  as  it  encapsulates
unpredictable system behavior.

A  failure  detector   is  specified  in  terms   of  two  properties:
\emph{completeness}        and       \emph{accuracy}~\cite{chandra96}.
Completeness  is  the  property  that describes  how  well  a  failure
detector will perceive  real failures, while accuracy  is the property
that describes how well  it will avoid mistakes~\cite{reynal05}, e.g.,
false  detections.  Failures  detectors will  behave  differently  and
support  other failures  models by  simply strengthening  or loosening
completeness and accuracy~\cite{guerraoui00}.

Of special interest for this work is the $\Omega$ failure detector,
originally presented in \cite{chandra96c}. The $\Omega$ failure
detector is a \emph{leader election}, a failure detector
that outputs a single trusted process. Formally, the $\Omega$ failure
detector is specified by the following properties~\cite{guerraoui00}:

\begin{itemize}
\item \textit{Eventual Accuracy}: There is a time after which every
correct process trusts some correct process.
\item \textit{Eventual Agreement}: There is a time after which no two
correct processes trust different processes.
\end{itemize}

This pair of properties ensures  every correct process will eventually
trust  the same  correct process.  The eventual  behavior means  it is
necessary  a  long  enough  period  of  synchrony  in  order  for  the
properties to be achieved.  The $\Omega$ failure detector is used as a
building   block   to   solve    more   complex   problems   such   as
consensus~\cite{lamport98} and atomic broadcast~\cite{chandra96c}.

It is  rather easy to  create a  leader election   using regular
failure detection. Let each process in a distributed system use a
failure detector to  monitor every other process and create  a set $C$
of processes it believes to be correct. Eventually the set $C$ will be
the same in all correct processes and  the leader can be chosen as the
process in $C$ with the smallest  pid. However, this reduction is very
costly  in terms  of  heartbeat messages,  as it  requires  $N^2 -  N$
heartbeat  messages  for  each  $\eta$, one  for  each  unidirectional
communication link.   Moreover, this reduction doesn't  provide a very
useful  property: \emph{leader  stability}.  Leader  stability is  the
ability of the failure detector to  never remove the leadership from a
correct  process  because  another  process (with  a  lower  pid,  for
instance) begins  to be  trusted~\cite{malkhi05}. Leader  stability is
very useful  for Paxos,  for example, because  a leadership  change in
this algorithm can be a costly operation~\cite{vieira14}.

\subsection{QoS of Failure Detection with Crash-Stop Failures}
\label{sec:chen}

A set  of failure detectors  with a model  for quality of  service was
first  proposed by  Chen et   al.~\cite{chen02}.  In  that work,  the
authors presented the concept of  QoS for failure detectors, alongside
quantitative  metrics to  measure it.   They also  presented a  set of
failure detectors  that were designed to  have a precise model  of the
behavior of these metrics.  Roughly speaking, quality of service means
the  failure  detector  is   configured  to  meet  strict  application
requirements  and  work  accordingly   to  the  network  probabilistic
behavior.  

The  failure  detector  proposed  by  Chen  et   al.   has  two  main
components:  an \textit{estimator}  and a  \textit{configurator}.  The
estimator  is  responsible  for analyzing  the  network  probabilistic
behavior  in  terms  of  message   losses  and  message  delays.   The
configurator is responsible for generating a suitable failure detector
based  on the  behavior  observed  by the  estimator  and  on the  QoS
requirements provided  by the application developer.   The application
developer must specify the application  requirements in terms of three
metrics:

\begin{itemize}
\item Detection time ($T_D$): This measures how much time elapses from
  the  occurrence of  a crash  and it  being detected  by the  failure
  detector;
\item  Mistake  recurrence time  ($T_{MR}$):  This  measures the  time
  between two consecutives mistakes made by the failure detector;
\item  Mistake  duration ($T_M$):  This  measures  how much  time  the
  failure detector takes to correct itself once it has made a mistake.
\end {itemize}

The result is a customized failure detector in which both heartbeat
inter-sending  interval  $\eta$ and  the  safety  margin $\alpha$  are
shaped  to meet  the  given  constraints, when  it  is feasible.  Once
configured, a failure detector is ready  to work as usual: a monitored
process $p$  sends every $\eta$  time units  a heartbeat message  to a
monitor process $q$, which will wait for those messages for a specific
time, plus a safety margin $\alpha$.

In  this paper  we  will discuss  only one  of  the failure  detectors
proposed by Chen et al.,  namely the \textit{New Failure Detector with
  Expected Arrival Time Calculation  (NFD-E)}.  The NFD-E algorithm is
the most suitable for partially synchronous distributed systems as it assumes
no clock synchrony among the  processes and estimates the arrival time
of  future heartbeats.   This  algorithm  is particularly  interesting
because its  QoS model is  based only on  the variance of  the message
propagation delays and not on  the unknown delays themselves.  However
there is  some limitations  about the  required assumptions  needed to
this algorithm work  properly.  It assumes a  crash-stop failure model
and, although there is no  assumption about synchrony among processes,
it  is necessary  that  each  process has  access  to  a local  clock.
Furthermore, it  is assumed there  is no  clock drift between  any two
local clocks.

The NFD-E algorithm works as follows:  for all $i \geq 1$, a monitored
process $p$ sends at time $i  \cdot \eta$ a heartbeat message $m_i$ to
a monitor process $q$.  Also for all $i \geq 1$, process $q$ waits for
message      $m_i$      until     a      \textit{freshness      point}
$\tau_i = EA_i + \alpha $. If no message with a label equals or grater
than $i$  is received by  $q$ until  $\tau_i$ expires, i.e.,  no fresh
message arrives  before its timeout,  then $q$ starts  suspecting $p$.
To   estimate  each   arrival  time   $EA_i$,  process   $q$  uses   a
\emph{predictor} based on the set  of $n$ previously received messages
from  $p$.   Let $m'_1,...,m'_n$  be  the  messages received  by  $q$,
$s_1,  ...   ,  s_n$  be   the  sequence  numbers  of  such  messages,
$A'_1,...,A'_n$ be their  receipt time according to  $q$'s local clock
and  $\ell$ the  largest  sequence number  among all  ${s_1,...,s_n}$.
Then $EA_{\ell+1}$ is estimated by~\cite{chen02}:

\begin{equation} \label{eq:estimateEA}
    EA_{\ell + 1} \approx \frac{1}{n} \left( \sum^n_{i=1}{A'_i - \eta
s_i} \right) + (\ell + 1) \eta
\end{equation}

The predictor  that calculates  $EA_i$ actually estimates  the average
message delay for the last $n$  messages, shifting backward in time by
$\eta \cdot  s_i$ each $A'_i$. By  adding $\alpha$ to estimation  made by
the predictor, the  failure detector is able to  absorb some variation
in message  delay. A  study made  by~\cite{nunes04} observed  that the
predictor used  in the NFD-E algorithm  is not the more  accurate, but
the  use of  a  constant  safety margin  $\alpha$  allows the  failure
detector to achieve a good QoS.  Hence the predictor used by the NFD-E
algorithm plays a more important role  in QoS achievement than the one
originally observed  by its authors, being  responsible for accounting
for  the   uncertainty  in  message  propagation   delays  typical  of
partially synchronous system.

The full algorithm as presented by its authors in~\cite{chen02} is
shown in Algorithm~\ref{alg_nfd-e}.

\begin{algorithm}
\caption{NFD-E Algorithm}\label{alg_nfd-e}
\begin{algorithmic}[1]
\Procedure {SendHeartbeat}{}
	\Comment {Procedure exclusive for $p$, using $p$'s local clock.}
	\For {all $i \geq 1$, at time $i \cdot \eta$} send heartbeat $m_i$ to
$q$
	\EndFor
\EndProcedure
\State
\Procedure{Initialization}{}
	\Comment {Procedures exclusives for $q$, using $q$'s local clock}
	\State {$\tau_0 \gets 0$}
	\State {$\ell \gets -1$}
\EndProcedure
\State
\Event {$ \tau_{\ell + 1} = now() $}
	\State $output \gets Suspect$
\EndEvent
\Event {$ \text{receives message }  m_j \text{ at time } t $}
	\If {$ j > \ell$}
		\State $\ell \gets j$
		\State $\tau_{\ell+1} \gets EA_{\ell + 1} + \alpha$ 
		\If {$t < \tau_{\ell+1}$} 
			\State $output \gets Trust$ 
		\EndIf
	\EndIf
\EndEvent
\State
\end{algorithmic}
\end{algorithm}

\subsection{QoS of Failure Detection with Crash-Recovery Failures}
\label{sec:ma}

In~\cite{ma010}  the  authors  analyze  the  QoS  of  the  synchronous
\emph{New Failure Detector with Synchronized Clocks (NFD-S)} algorithm
found in~\cite{chen02}.   This algorithm assumes a  synchronous system
and Ma et al. assume some  sort of time synchronization to be present,
such as the NTP protocol.

The expanded  model outlined in~\cite{ma010} presented  additional QoS
metrics that complement those in~\cite{chen02} listed
in Section~\ref{sec:chen}:

\begin{itemize}
\item Query Accuracy Probability  ($P_A$): it measures the probability
  that,  at any  arbitrary  time when  queried,  the failure  detector
  correctly indicates the state of the monitored process.
\item Recovery  Detection Time  ($T_{DR}$): it  measures the  time the
  failure detector perceives a recovery at the monitored process.
\item Detected Failure Proportion ($R_{DF}$): it measures the ratio of
  the detected crashes over the actual crashes.
\end{itemize}
 
The configuration of the failure detector and parameter estimation for
a  desired  QoS  are  shown  in~\cite{ma010}.   While  an  interesting
expansion of  previous work,  \cite{ma010} conclude that  the proposed
algorithm and  QoS model  needs to be  enhanced, specifically  to deal
with  short failures  that  may not  be  detected.  Additionally,  the
experimental  results in  some  tests differs  substantially from  the
expected behavior  of the implemented failure  detector (specially for
the $R_{DF}$ QoS metric), reinforcing  the need for more research for
a failure detector in this system model. Because of these limitations,
in this paper we do not use the expanded set of metrics for a crash-recovery system, but instead we use the basic set proposed by~\cite{chen02}. 


\section{Algorithm} 

In  the  previous sections,  we  provided  the theoretical  background
needed to  understand key concepts  such as synchrony  models, failure
detectors with quality of service and leader election. In this section
we  present the \textit{New  Failure Detector as a Leader Election
  Algorithm (NFD-L)} leader election  algorithm, which is a derivative
of        NFD-E        algorithm~\cite{chen02},        shown        in
Algorithm~\ref{our_leader_election}.   NFD-L  is an  efficient  leader
election  with stability  that  only requires  a  single message  each
$\eta$.  More importantly, NFD-L is designed to work on the
crash-recovery system  model with  no clock synchrony  among processes
and  requiring  only  a  single stable  memory  write  during  process
initialization.

\begin{algorithm}
\caption{NFD-L Algorithm}\label{our_leader_election}
\begin{algorithmic}[1]
\Procedure{Initialization}{}
		\State {$ leader \gets \bot $}
		\State {$ \mathit{self.uptime}    \gets 0 $}
		\State $\textit retrieve( self.zerotime )$
		\If {$ self.zerotime = \bot  $}
			\State $\textit{self.zerotime} \gets \textit{now()}$
			\State $\textit{store(self.zerotime)}$
		\EndIf
	\State {$ i \gets \frac{now() - self.zerotime}{\eta} $}
\EndProcedure
\Procedure {SendHeartbeat}{}
	\If {$self.pid = leader$}
		\For {all $i \geq 1$, at time $i \cdot \eta$} 
			\State {send heartbeat $m_i$ to all processes}
			\State {$ self.uptime \gets self.uptime + 1$}
		\EndFor
	\EndIf
\EndProcedure
\State
\Event {$ \text{receives message }  m_j \text{ at time } t $}
	\If {$ sender(m_j) = leader $}
		\If {$j > \ell$}
			\State $\ell \gets j$
			\State $\tau_{\ell+1} \gets EA_{\ell + 1} + \alpha$
		\EndIf

	\Else
		\If {$ uptime(sender(m_j)) > uptime(leader)$} 
			\State $\ell \gets j$
			\State $leader \gets sender(m_j)$
			\State $output \gets leader $
		\Else
			\If{$uptime(sender(m_j)) = uptime(leader)$}
				\Comment{Tiebreaker}
				\If {$ pid(sender(m_j)) > pid(leader)  $}
					\State $\ell \gets j$
					\State $leader \gets sender(m_j)$
					\State $output \gets leader$
				\EndIf
			\EndIf
		\EndIf
	\EndIf
\EndEvent
\State
\Event {$ \tau_{\ell + 1} = now() $}
	\State $leader \gets self.pid$
	\State $output \gets leader$
\EndEvent

\end{algorithmic}
\end{algorithm}


\subsection{A Failure Detector with QoS as a Leader Election}

A leader election is a failure  detector that outputs a single trusted
process.  The output of NFD-L is the $pid$ of the trusted $leader$. It
is  possible  that each  process  of  the  distributed system  sees  a
different process as a $leader$ at  the same time, but when the system
behaves synchronously for a sufficient  amount of time, eventually all
correct processes will agree on the  same $leader$. To achieve this we
combine      the     general      principle      of     the      bully
algorithm~\cite{garcia-molina82} with the failure detecting properties
of the NFD-E algorithm, using process  uptime as a priority measure to
ensure a basic level of stability.

Each process  starts (or recovers)  knowing nothing about  the current
leader.  If  it receives  no message  from a  leader process,  it then
assumes it is the leader and starts sending heartbeat messages.  A
process loses the leadership when it receives a message from a process
with greater priority: greater $uptime$  or greater $pid$ in the event
of a tie on $uptime$.  When  a process recognizes another process as a
leader, it  stops sending heartbeats and  starts monitoring heartbeats
sent by  the leader,  in a  behavior similar to  NFD-E. Thus,  at this
moment, only  a process  acting as a  leader sends  periodic heartbeat
messages to  all other  processes that will  behave as  monitors.  The
leader election  stabilizes when each  correct process receives
the message of a single process with high enough priority.

In  steady  state operation  and  assuming  a broadcast  communication
medium the cost of NFD-L will  be equivalent to single instance of the
NFD-E algorithm,  compared to  the $N^2  - N$  instances of  the naive
reduction. Furthermore, in the  fail-free case after stabilization the
QoS  of NFD-L  will be  equivalent to  the  QoS of  NFD-E as  it is  a
generalization of this algorithm.

\subsection{Dealing with Crashes and Recoveries}

Each heartbeat message $m$ sent by  a monitored process $p$ carries an
incremental  sequence number  $i$ used  by  a monitor  process $q$  to
determine if  an incoming heartbeat  is still fresh, thus  trusting or
suspecting $p$ to remain as leader.  After recovering from a crash, it
is important that process $p$  starts sending correct sequence numbers
to $q$,  i.e., process  $p$ should  remember which  one was  the exact
sequence  number  $i$ it  sent  before  it  crashed.  If  process  $p$
``forgets''  its  sequence number  and  starts  sending $i$  from  the
initial value, it will not be trusted by process $q$ until $i$ reaches
the value it held before process $p$ crashed.

The naive solution  to remember the correct sequence number  $i$ to be
used after a  crash, is to write on persistent  storage $i$ every time
process $p$ sends a heartbeat  message $m_i$.  However, while correct,
this solution isn't efficient regarding the number of writes to costly
stable memory.  We propose a cheaper solution that uses a single write
operation once  a process is initialized  and to a read  operation per
process recovery.

Our solution takes advantage of  the assumption that every process has
access to a  local clock and that the $i$-th  heartbeat message should
be sent at time  $i \cdot \eta$.  It works as  follows: when a process
starts for the  very first time or it recovers  from a previous crash,
it checks if it wrote on  persistent storage a timestamp of a previous
initialization. If  there is  not a  previous timestamp  on persistent
storage, the process  then writes the current  timestamp on persistent
storage.  This  way, this  value will  be read  and constant  at every
single recovery of that process. Using this value a process calculates
what  the current  message label  $i$ is  supposed to  be, based  on a
function of the elapsed time since  the first startup and $\eta$ (Line
8).  It  ensures that  every time  a process  starts or  recovers, the
message   label  $i$   will   not  violate   the  properties   defined
in~\cite{chen02}, i.e., every message label $i$ sent by a process will
be greater than the previous label $i-1$. The only requirement is that
the local clock keeps increasing with no drift during crashes.

\section{QoS Analysis}\label{sec_results} 

In  this   section  we   present  the  QoS   analysis  of   the  NFD-L
algorithm. The QoS  analysis is made using the  three metrics proposed
in~\cite{chen02}, $T_D$, $T_{MR}$ and $T_M$ (Section~\ref{sec:chen}),
with the  addition of the $T_{DR}$  metric proposed in~\cite{ma010}  for
the  crash-recovery   failure  model   (Section~\ref{sec:ma}).   These
metrics  will   be  grouped in  failure   detector  accuracy  metrics
($T_{MR}$,  $T_M$)   and  failure   detector  speed   metrics  ($T_D$,
$T_{DR}$).  The experiments have the objective of assessing if the QoS
of  the proposed  failure detector  is within  the bounds  achieved by
Chen's algorithm.

\subsection{Enviroment and configuration of the leader election}

These experiments  ran at the Maritaca  computational cluster, located
at Universidade Federal de São Carlos,  Sorocaba, São Paulo. We used 5
identical nodes, each one equipped with an 8 cores/16 threads processor
and  16 GB  of volatile  memory. Each  node used  a hard  disk as  its
persistent  storage and  all nodes  were interconnected  by a  Gigabit
Ethernet link. 

The NFD-L  algorithm by design  selects a node  as the leader  and the
other  four nodes  behave as  monitor processes,  receiving heartbeats
sent by the  leader.  We monitored the leader election  output of each
monitor process  to assess  the QoS metrics.   To generate  a constant
load on the network and on the  system, we used the leader election as
a     service     provider     for    the     Treplica     replication
framework~\cite{vieira08a}.   The  replication  rate was  set  to  the
saturation point  of the application  with this setup, which  was 2700
operations per second on average.

To  generate  the  parameters  $\eta$  and  $\alpha$  for  the  leader
election, we used the configurator described by~\cite{chen02} during a
  period  when the  network  was  under high  load  due  to the  test
application.  We then measured  the network probabilistic behavior for
1 hour, repeating it  3 times in total in different  hours of the day.
The  observed  message loss  probability  $p_L  = 0.0175917$  and  the
observed message delay  variance $V(D) = 25.3356$ were  then used with
the following QoS  metrics $T_D = 1000$ ms, $T_{MR}  = 3600000$ ms and
$T_M = 1000$ ms to obtain $\eta = 330$ ms and $\alpha = 670$ ms.

We made  two set of experiments  in order to respectively  observe the
accuracy  and the  speed of  the failure  detector.  The  accuracy was
measured  by counting  the  average mistake  rate  ($T_{MR}$) and  the
mistake duration ($T_M$)  for each monitor process during 6  runs of 1
hour each. During the experiment the leader process did not fail, thus
all  suspicions were  mistakes  that were  eventually corrected.   Let
$t_{m0},...,t_{mn}$  be  the timestamps  of  the  mistakes made  by  a
process and $t_{r0},...,t_{rn}$ be the timestamps of the correction of
those mistakes, then we computed the mistake rate of that process as:
\[\ \frac{1}{T_{MR}} = \frac{1}{ \frac{1}{n}
\left(\sum^n_{i=1}{(t_{mi} - t_{mi-1}) } \right)   } \] 

The average mistake duration of a process was computed as the average
time taken to correct a mistake: 

\[\ T_M = \frac{1}{n} \left(\sum^n_{i=0}{( t_{ri} - t_{mi}  )} \right) \] 

The speed of  the leader election algorithm was measured  by the crash
detection time ($T_D$) and by  the recovery detection time ($T_{DR}$).
To be able to measure $T_{DR}$, we  have randomly selected a node as a
high priority leader, breaking the  specification of NFD-L in such way
that this leader would be  immediately reelected after a recovery.  We
then  injected a  single  crash  in the  leader  process  and then  we
recovered it,  measuring the  time elapsed for each  process to
detect the  crash and, afterwards,  the time elapsed for  each process
detect  the   leader  recovery.   This  experiment   consisted  of  10
repetitions of a crash-recovery cycle, with 60 seconds between a crash
and the following  recovery.  Let $t_c$ be the time  of a leader crash
and $t_d$  the time of detection  of that crash by  a monitor process,
the detection time $T_D$ was  obtained by calculating the difference of
times:
\[\ T_D = t_d - t_c \]

In  a very  similar way,  the recovery  detection time  was calculated
considering the  recovery time  $t_r$ and the  detection time  of that
recovery $t_{dr}$ by a process:
\[\ T_{DR} =  t_{dr} - t_r  \]

For this experiment  we used the NTP protocol at  the local network to
synchronize the clocks of the nodes of the cluster.  This procedure is
not necessary for  the algorithm to work properly, but  it made easier
to calculate  time differences  between the nodes.   By using  the NTP
protocol we also  observed that the average message  delay between the
nodes were a  thousand of times smaller than  the message intersending
interval  $\eta$  intended  to  be  used in  our  experiments,  so  we
considered the message delay negligible.

\subsection{Analysis}

We  plotted the  QoS  metrics  observed in  the  experimental runs  in
boxplots.   For each  run  the  data points  represent  the metric  as
measured by each  one of the four monitor process  with respect to the
current state  of the actual  leader.  This amounts  to a total  of 24
points  for the  experiments  assessing the  accuracy  of the  failure
detector and 40 points for the  experiments assessing the speed of the
failure  detector.  The  expected  value  of the  QoS  metric used  to
configure  the NFD-L  algorithm is  shown by  the traced  line in  all
plots. Numeric values  for the data points represented  in the figures
are shown  in Table~\ref{table:boxplot_values}, listing the  values of
the 1st, 2nd and 3rd quartiles.

\begin{figure}[htp]
    \centering 
    \includegraphics[width=13cm]{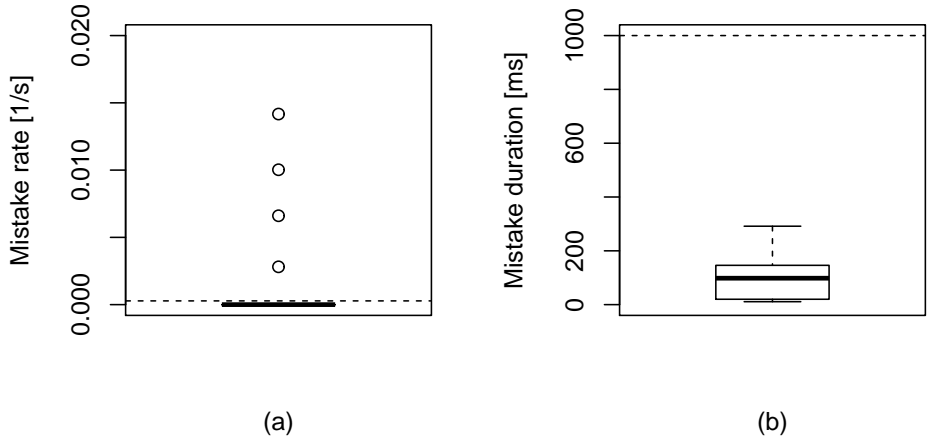}
    \caption{(a) The  mistake rate  $1 / T_{MR}$  and (b)  the mistake
      duration $T_M$. The traced  line in the plots represents the
      upper bound QoS requirements for the leader election.}
    \label{fig:accuracy} 
\end{figure}

Figure~\ref{fig:accuracy}(a)  shows the  observed  mistake rate.   The
majority of the processes (the darker line at the bottom of the image)
made  no   mistakes,  successfully  achieving  the   required  QoS  of
$T_{MR} = 3600000$ ms. However, there were outlier processes that did
not achieve the  required QoS. Among these, there is  a single outlier
that is not shown on figure,  with a mistake rate of 33.34 mistakes/s.
These  outliers  consist of  a  couple  of  mistakes  in an  one  hour
experiment, negatively affecting the  average mistake recurrence time.
In fact, the way the metric is computed (as proposed in~\cite{chen02})
increases  the impact  of each  of the  low number  of mistakes.   For
example,  the 33.34  mistakes/s outlier  was  created by  one pair  of
mistakes  30 ms  apart.   The  Figure~\ref{fig:accuracy}(b) shows  the
observed mistake duration.   The processes that made  no mistakes were
disconsidered. All  the observed  processes did  meet the  required QoS
$T_M = 1000$ ms.

\begin{figure}[htp]
\centering
\includegraphics[width=13cm]{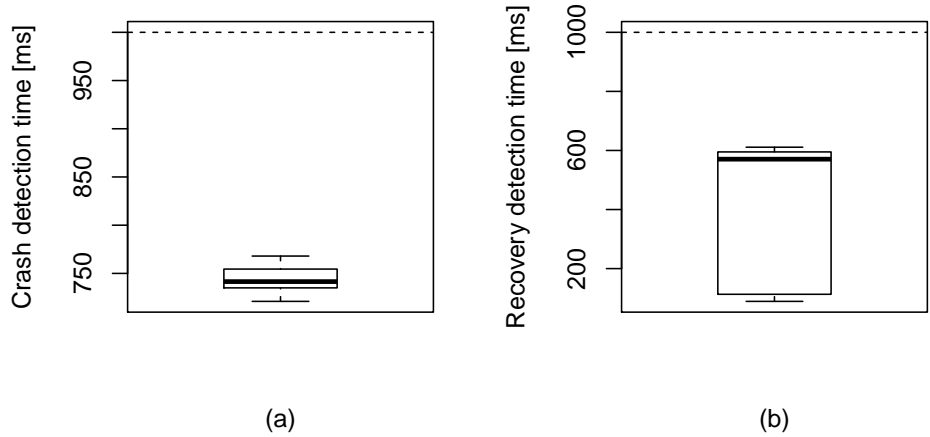}
\caption{(a) The leader crash detection time $T_D$ and (b) the leader
recovery detection time $T_{DR}$ 
observed by each monitor process.
The traced line represents the upper bound  QoS requirements for the
leader election.}
\label{fig:speed}
\end{figure}

The crash  detection time  is shown in  Figure~\ref{fig:speed}(a). All
the  crashes  were detected  within  the  required  QoS as  the  crash
detection   is   bounded   to   $T_D    =   \eta   +   \alpha$.    The
Figure~\ref{fig:accuracy}(b)   shows  the   recovery  detection   time
$T_{DR}$ observed  by the  monitor processes.   The speed  metrics had
less variation than  the accuracy ones.  It suggests  that accuracy is
more sensitive to process  dependability than to network probabilistic
behavior.   As we  assessed the  QoS metrics  of the  failure detector
integrated in  a loaded application,  pauses in the processing  of the
application created loss  and delay of messages more  intense than the
ones created by network probabilistic behavior alone. In a sense, this
amounts  to periods  of omission  faults where  the processes  stopped
processing.  We  conjecture that  a QoS  model for  the crash-recovery
process model should be able to absorb this process behavior.

\begin{table}[htbp]
\centering
\begin{tabular}{l*{4}{c}r}\label{table:boxplot_values}
QoS Metric & 1st. Quartile & Median & 3rd. Quartile & Upper bound QoS\\
\hline
$1/ T_{MR}$  (Figure~\ref{fig:accuracy}(a))      &  0 1/s  & 0
1/s    & 0 1/s     &  $0.278 \cdot 10^{-3} $ 1/s\\
$T_M$   (Figure~\ref{fig:accuracy}(b))            & 20 ms &
98.223 ms & 146.03 ms &  1000 ms\\
$T_D$ (Figure~\ref{fig:speed}(a))             &  735 ms& 741.5
ms & 754.25 ms &  1000 ms\\
$T_{DR}$ (Figure~\ref{fig:speed}(b))          & 113 ms & 570.5
ms & 595  ms  &  1000 ms\\
\end{tabular}
\caption{Summary of plot values}
\label{table:boxplot_values}
\end{table}

\section{Conclusion}

In  this  paper  we  presented  the  NFD-L  leader  election  for  the
asynchronous  crash-recovery  failure  model.  This  algorithm  is  an
extension of  the NFD-E~\cite{chen02} algorithm  adapted to work  as a
leader election that is efficient in terms of its use of stable memory
and message exchanges.  The NFD-L  algorithm used two novel techniques
to achieve its  efficiency: (i) instead of using a  counter of crashes
as usual~\cite{martin09},  it uses a  counter of uptime  to prioritize
stable processes and  (ii) it uses a single write  in stable memory to
create a sequence of heartbeats that isn't interrupted by crashes, but
only  appears to  be ``paused'',  reducing crash  failures to  message
omission failures.

We analyzed the performance of the NFD-L algorithm on a real system to
verify  how  well  it  met  the  given  QoS,  configured  as  proposed
in~\cite{chen02}. We  were able  to achieve the  desired QoS,  but the
accuracy  QoS metrics  suffers  for repetitive  short period  mistakes
caused by  the application.   Our results suggest  that accuracy  of a
failure  detector is  dependent on  process dependability  besides the
effects of network probabilistic  behavior. In particular, application
overload spikes,  lock contention  and other performance  problems can
severely  interfere in  the generation  and processing  of heartbeats.
Thus, further  enhancements of the  QoS model  are needed to  make the
algorithm suitable to an  environment where the processes dependability
is  a   main  concern.   We   believe  these  changes  must   take  in
consideration  crash-recovery  parameters  such   as  the  ones  found
in~\cite{ma010}  or  be made  more  application  specific as  proposed
in~\cite{hayashibara04}.

\section*{Acknowledgments}

We would  like to  thank Priscila  Aiko Someda  Dias for  her valuable
support during the  statistical analysis of this  work.  This research
was funded by the  Brazilian \textsl{Coordenação de Aperfeiçoamento de
  Pessoal     de     Nível     Superior     (CAPES)}     under     the
\textsl{Pró-Equipamentos}   program    (Edital   25/2011)    and   the
\textsl{Demanda Social} program.

\bibliographystyle{apalike}
\bibliography{qos.leader.election}

\end{document}